# On the Blasius correlation for friction factors

Trinh, Khanh Tuoc
Institute of Food Nutrition and Human Health
Massey University, New Zealand
*K.T.Trinh@massey.ac.nz*

---

## Abstract

The Blasius empirical correlation for turbulent pipe friction factors is derived from first principles and extended to non-Newtonian power law fluids. Two alternative formulations are obtained that both correlate well with the experimental measurements of Dodge, Bogue and Yoo.

## 1 Introduction

In a previous paper (Trinh, 2010a) logarithmic correlations for turbulent pipe flow of Ostwald de Waele fluids have been derived from a master curve for instantaneous friction factors (Trinh, 2009a). Blasius (1913) has also proposed an empirical power law correlation

$$f = \frac{2\tau_w}{\rho V^2} = \frac{0.079}{\mathrm{Re}^{1/4}} \tag{1}$$

where $\mathrm{Re} = (DV\rho/\mu)$ is the Reynolds number, $D$ the pipe diameter, $\rho$ the fluid density, $V$ the average velocity, $\mu$ the fluid viscosity and $\tau_w$ the time-averaged wall shear stress. Nikuradse (1932) proposed that the Blasius correlation corresponds with a power law representation of the velocity profile

$$U^+ = A\left(y^+\right)^p \tag{2}$$

where the exponent p is a function of the Reynolds number, $y$ is the radial distance from the wall, $U^+ = U/u_*$ and $y^+ = yu_*\rho/\mu = yu_*/\nu$ have been normalised with the friction velocity $u_* = \sqrt{\tau_w/\rho}$. An alternative form of the power law profile is

$$\frac{U}{U_m} = \left(\frac{y}{R}\right)^p \tag{3}$$

Nikuradse (op.cit.) and Prandtl (1935) also proposed an alternative representation of the velocity profile

$$U^+ = 2.5 \ln y^+ + 5.5 \tag{4}$$

Where At the wall Prandtl postulated the existence of a laminar sub-layer described by

$$U^+ = y^+ \tag{5}$$

which applies up to $y^+ \leq 5$. There is a long and on-going debate as to whether the power law represented by equation (2) or the so called logarithmic law of the wall, log law in short, best describes experimental measurements of velocity profiles in turbulent flow e.g. (McKeon et al., 2004).

Dodge and Metzner (1959) have extended the Blasius correlation to purely viscous non-Newtonian fluids

$$f = \frac{\alpha}{\mathrm{Re}_g^{\beta}} \tag{6}$$

where

$$\mathrm{Re}_g = \frac{D^n V^{2-n} \rho}{K 8^{n-1} \left(\frac{3n+1}{4n}\right)^n} \tag{7}$$

Is called the Metzner-Reed (1955)generalised Reynolds number and the rheological parameters $K', n'$ are defined by the laminar equation

$$\tau_w = K' \left(\frac{8V}{D}\right)^{n'} = K \dot{\gamma}_w^n \tag{8}$$

For Ostwald de Waele power law fluids

$$n' = n$$
$$K' = K\left(\frac{3n+1}{4n}\right) \quad (9)$$

In this paper, we will derive the Blasius and Dodge-Metzner empirical equations from theoretical considerations.

## 2 Theory

There are two possible ways to derive a power law relationship for non-Newtonian fluids which must of course also apply to Newtonian fluids when $n = 1$.

## 2.1 Extension of the Blasius empirical correlation

In this approach we begin with the observation that all pipe friction factors can be described by a unique master curve when expressed in terms of the critical instantaneous shear stress (critical shear stress for short) at the point of bursting in the wall layer process (Trinh, 2009a). The Blasius correlation can then be written as:

$$f_e = \frac{0.04}{Re_e^{1/4}} \quad (10)$$

Where

$$f_e = \frac{2\tau_e}{\rho V^2} \quad (11)$$

$$Re_e = \frac{DV}{\nu_e} = \frac{DV}{K^{1/n}\tau_e^{(n-1)/n}} \quad (12)$$

are the friction factor and Reynolds number based on the critical instantaneous wall shear stress $\tau_e$. For high Reynolds numbers, the wall layer is thin (Trinh, 2009b) and the radius of curvature neglected. Then

$$\tau_w = (n+1)\tau_e \quad (13)$$

$$\mathrm{Re}_e = \left(\mathrm{Re}_g f^{1-n'}\right)^{\frac{1}{n'}} 2^{\frac{5(n'-1)}{n'}} \left(\frac{(3n'+1)}{4n'}\right) \left(\frac{n'+1}{2}\right)^{\frac{n'-1}{n'}} \tag{14}$$

Substituting equations (13) and (14) into (10) gives

$$f = \frac{\alpha}{\mathrm{Re}_g^{1/(3n'+1)}} \tag{15}$$

Where

$$\alpha = \left(0.079 \frac{n+1}{2}\right)^{\frac{4n}{3n+1}} (2)^{\frac{5(1-n)}{3n+1}} \left(\frac{4n}{3n+1}\right)^{\frac{n}{3n+1}} \left(\frac{2}{n+1}\right)^{\frac{n-1}{3n+1}} \tag{16}$$

## 2.2 Matching the power law profile with the wall layer

The most glaring weakness of the power law velocity profile is its infinite gradient at the wall which is physically unrealistic. The problem can be circumvented by observing that the ejections of wall fluids that begin outside the wall layer at a time averaged distance $y^+ = 30$ which is the time-averaged value of the wall layer thickness $\delta_v$. We force equation (3) through the edge of the wall layer

$$\frac{U_v^+}{U_m^+} = \frac{\phi U_v^+}{V^+} = \left(\frac{\delta_v^+}{R^+}\right)^p \tag{17}$$

We use the value $p = 1/7$ to keep in the same range of Reynolds number as the Blasius equation. Substituting for

$$V^+ = \sqrt{\frac{2}{f}} \tag{18}$$

and

$$R^+ = \mathrm{Re}_g^{1/n} f^{\frac{1}{n}-\frac{1}{2}} \left(\frac{3n+1}{4n}\right) 2^{\frac{5n-8}{2n}} \tag{19}$$

and rearranging

$$f = \frac{\alpha'}{\mathrm{Re}_g^{1/(3n+1)}} \tag{20}$$

The power index for the Reynolds number in equations (15) and (20) are the same which is to be expected. But now

$$\alpha' = \frac{2^{\frac{n+4}{(3n+1)}} \phi^{-\frac{7n}{(3n+1)}} \delta_v^{+\frac{n}{3n+1}}}{U_v^{+\frac{7n}{(3n+1)}} \left(\frac{3n+1}{4n}\right)^{\frac{n}{3n+1}}} \tag{21}$$

The ratio $\phi$ can be calculated by integrating equation (3)

$$\phi = \frac{2}{(p+1)(p+2)} = 0.817 \tag{22}$$

The problem boils down to the an estimate of $U_v^+$ *and* $\delta_v^+$. For a Newtonian fluid, $n = 1$, these values are well known (Trinh, 2009b, Trinh, 2010b): $\delta_v^+ = 64.8$, $U_v^+ = 15.58$. Then $\alpha' = 0.0787$.

For large Reynolds numbers, the wall layer thickness is thin compared to the pipe radius $\delta_v^+ << R^+$, the radius of curvature can be neglected and the wall layer can be analysed like a viscous sub-layer on a flat plate (Trinh, 2009a). Then the wall layer thickness and velocity are given as

$$\delta_v^+ = 64.8 \frac{(n+1)^{\frac{2-n}{2n}}}{\sqrt{2}} = 45.85(n+1)^{\frac{2-n}{2n}} \tag{23}$$

$$U_v^+ = \frac{\delta_v^+}{2.08(n+1)^{1/n}} = 45.82 \frac{(n+1)^{\frac{2-n}{2n}}}{2.08(n+1)^{1/n}} \tag{24}$$

Then

$$\alpha' = \frac{2^{\frac{n+4}{(3n+1)}} \phi^{-\frac{7n}{(3n+1)}} 45.82^{\frac{-6n}{3n+1}} (n+1) 2.08^{\frac{7n}{3n+1}}}{\left(\frac{3n+1}{4n}\right)^{\frac{n}{3n+1}}} \tag{25}$$

## 3 Results and discussion

The derivations presented here have been compared with the experimental of Dodge (1959), Bogue (1962) and Yoo (1974). The values of $\alpha$ estimated from equations (16) and (25) differ by approximately 1%. An example using equation (16) is shown in Figure 1

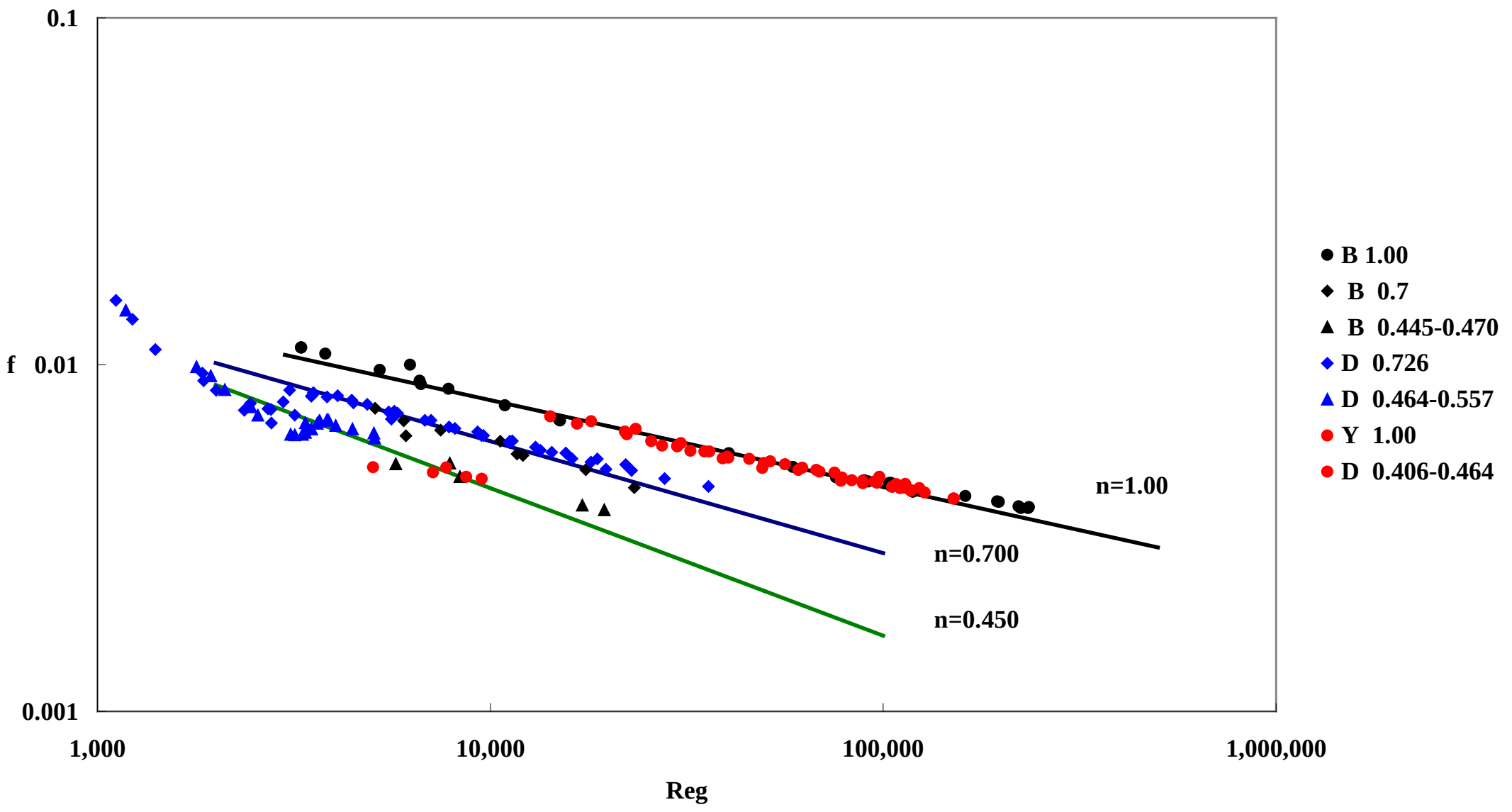

Figure 1 Comparison of predictions from equation (15 and 16) with measured friction factors. Data of Dodge (1959) Bogue (1962) and Yoo (1974).

Equations (15) and (20) both correlate 269 data points with a standard deviation of about 4.9% and a standard error of 0.03%. A comparison between the measured friction factors with the predictions by equation (15) is shown in Figure 2.

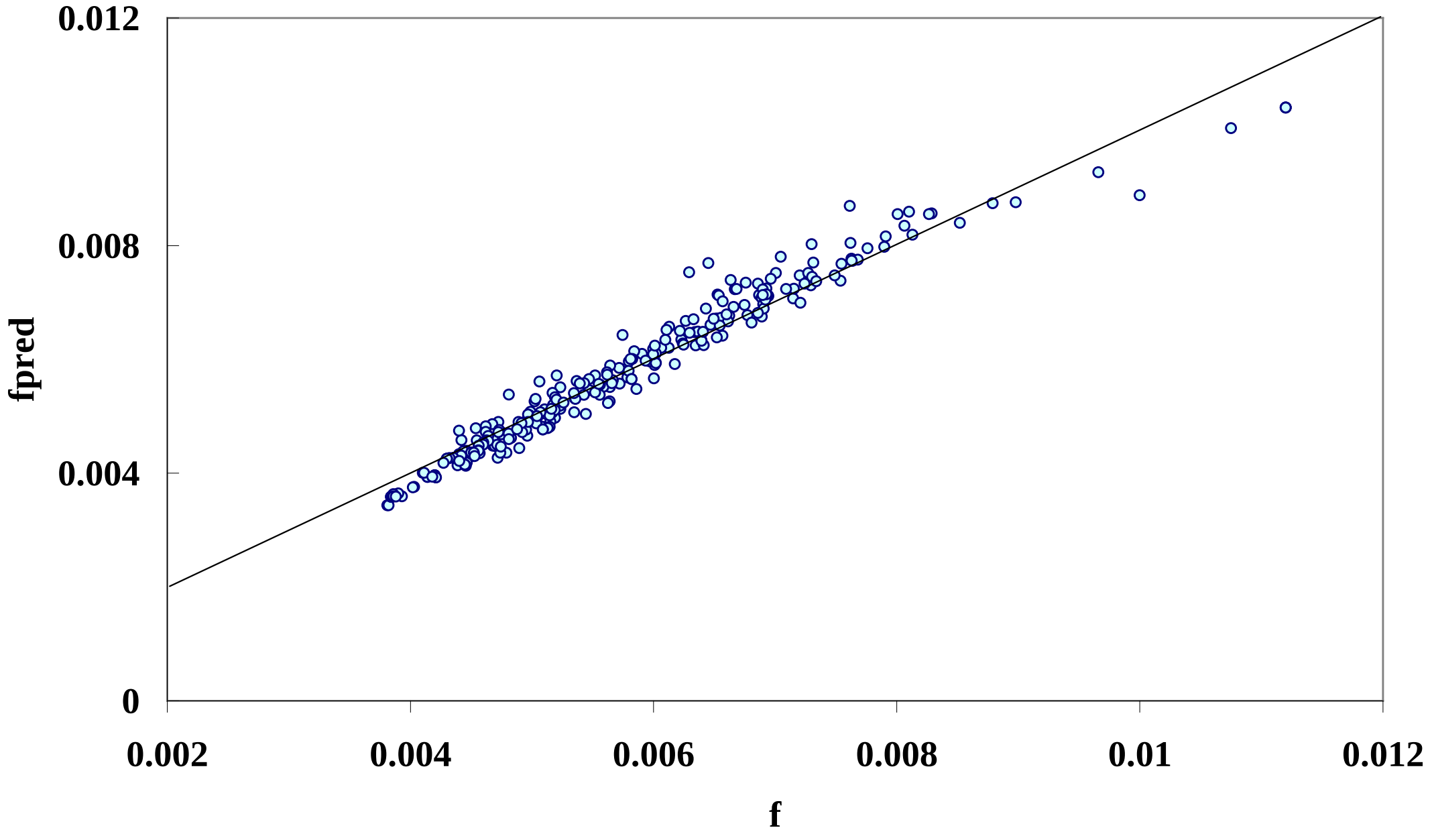

Figure 2  Comparison of measured friction factors and with calculation from equation (15). Same data as Figure 1.

However, Blasius type correlations tend to underestimate the experimental data for $\mathrm{Re}_g > 10^5$ and overestimate them for $\mathrm{Re}_g < 4000$. This is shown more clearly in Figure 3. Thus the low standard errors give a slightly distorted view of prediction accuracy.

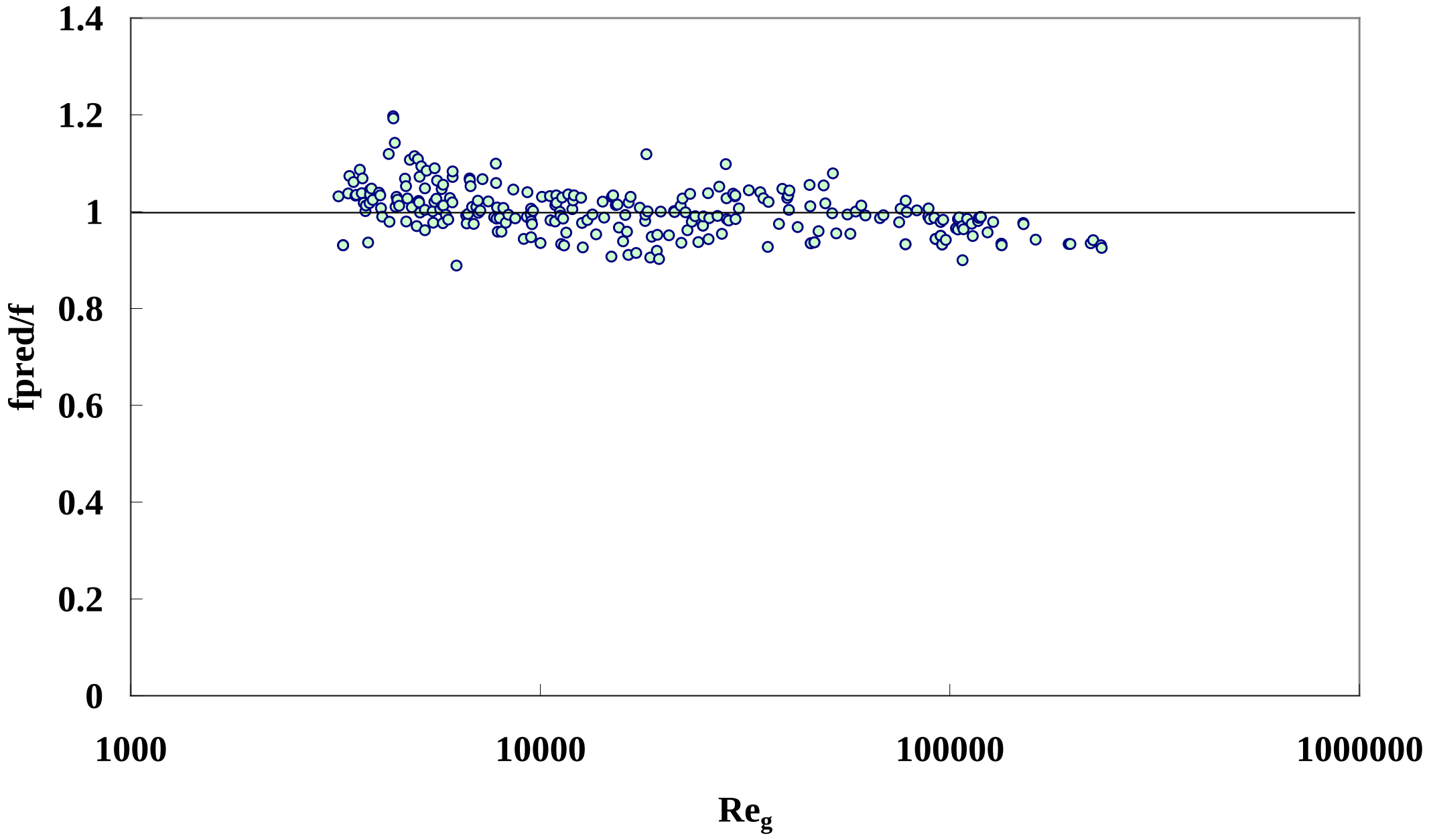

Figure 3  Effect of Reynolds number on the prediction of friction factors using equation (15).

While the predictions of friction factors from the extended Blasius correlation of Dodge-Metzner and in this work are similar, the dependence of $\alpha$ and $\beta$ on $n$ are quite different as seen in Figure 4.

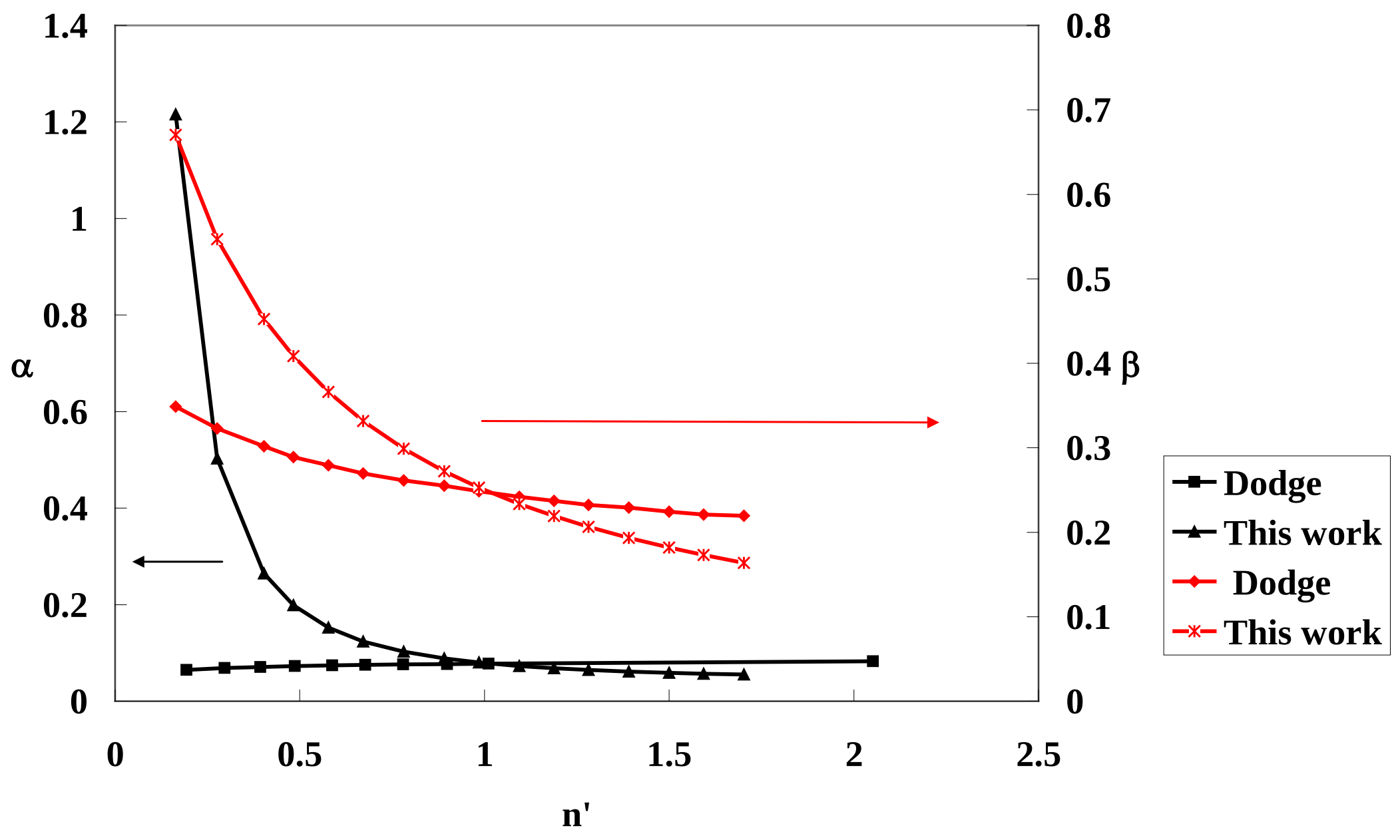

Figure 4 Comparison between experimental values of $\alpha, \beta$ of Dodge and equation (15).

The variations of $\alpha$ and $\beta$ in this work are much stronger, especially at small values of $n$. The relation between $p$ and $\beta$ can be formally established (Skelland, 1967, Trinh, 1992, Trinh, 2009b). Equation (15) can be solved for V as

$$\frac{2\tau_w}{\rho V^2} = \frac{\alpha}{\left[\dfrac{D^n V^{2-n} \rho}{K 8^{n-1} \left(\dfrac{3n+1}{4n}\right)^n}\right]^\beta} \tag{26}$$

Solving for $V$

$$V = A_1 \tau_w^{\frac{1}{2-\beta(2-n)}} R^{\frac{-n\beta}{2-\beta(2-n)}} \tag{27}$$

Where

$$A_1 = \alpha^{\frac{1}{\beta(2-n)-2}} 2^{\frac{2n\beta-3\beta-1/2}{2-\beta(2-n)}} \left(\frac{4n}{3n+1}\right)^{\frac{n\beta}{2-\beta(2-n)}} \rho^{\frac{\beta-1}{2-\beta(2-n)}} \tag{28}$$

Equation (3) is rearranged as

$$U = V\phi\left(\frac{y}{R}\right)^p \tag{29}$$

Substituting for $V$ from (27)

$$u = A_1 \phi y^p \tau_w^{\frac{1}{2-\beta(2-n)}} R^{-p-\frac{n\beta}{2-\beta(2-n)}} \tag{30}$$

Within the wall layer, the velocity distribution is independent of the radius (Trinh, 2009b, Skelland and Sampson, 1973) which corresponds to a zero exponent for $R$. Then

$$p = \frac{\beta n}{2-\beta(2-n)} \tag{31}$$

$$\beta = \frac{2p}{n+2p-np} \tag{32}$$

In the particular case when $p = 1/7$, equation (32) gives

$$\beta = \frac{1}{3n+1} \tag{33}$$

as shown in equations (15) and (20). In fact the index $p$ can be further expressed in terms of the Reynolds number by matching the logarithmic and power law correlations of the velocity profile (Trinh, 1992, Trinh, 2009b). This approach makes extends the range of application of Blasius type correlations to a larger range of Reynolds number. Details will be presented in a separate paper.

# 4 Conclusion

The Blasius empirical correlation has been successfully developed and extended to power law non-Newtonian fluids from theoretical considerations.

# 5 References

BLASIUS, P. R. H. 1913. Das Aehnlichkeitsgesetz bei Reibungsvorgangen in Flüssigkeiten. *Forschungsheft* 131**,** 1-41.

BOGUE, D. C. 1962. *Velocity profiles in turbulent non-Newtonian pipe flow, Ph.D. Thesis,* University of Delaware.

DODGE, D. W. 1959. *Turbulent flow of non-Newtonian fluids in smooth round tubes, Ph.D. Thesis,* US, University of Delaware.

DODGE, D. W. & METZNER, A. B. 1959. Turbulent Flow of Non-Newtonian Systems. *AICHE Journal,* 5**,** 189-204.

MCKEON, B. J., LI, J., JIANG, W., MORRISON, J. F. & SMITS, A. J. 2004. Further observations on the mean velocity distribution in fully developed pipe flow. *Journal of Fluid Mechanics,* 501**,** 135-147.

METZNER, A. B. & REED, J. C. 1955. Flow of Non-Newtonian Fluids - Correlation of the Laminar, Transition, and Turbulent-Flow Regions. *Aiche Journal,* 1**,** 434-440.

NIKURADSE, J. 1932. Gesetzmäßigkeit der turbulenten Strömung in glatten Rohren. *Forsch. Arb. Ing.-Wes. N0. 356.*

PRANDTL, L. 1935. The Mechanics of Viscous Fluids. *In:* W.F, D. (ed.) *Aerodynamic Theory III.* Berlin: Springer.

SKELLAND, A. H. P. 1967. *Non-Newtonian Flow and Heat transfer,* New York, John Wiley and Sons.

SKELLAND, A. H. P. & SAMPSON, R. L. 1973. Turbulent non-Newtonian boundary layers on a flat plate. *The Chemical Engineering Journal,* 6**,** 31-43.

TRINH, K. T. 1992. *Turbulent transport near the wall in Newtonian and non-Newtonian pipe flow, Ph.D. Thesis,* New Zealand, University of Canterbury.

TRINH, K. T. 2009a. The Instantaneous Wall Viscosity in Pipe Flow of Power Law Fluids: Case Study for a Theory of Turbulence in Time-Independent Non-Newtonian Fluids. *arXiv.org 0912.5249v1 [phys.fluid-dyn]* [Online].

TRINH, K. T. 2009b. A Theory Of Turbulence Part I: Towards Solutions Of The Navier-Stokes Equations. *arXiv.org 0910.2072v1 [physics.flu.dyn.]* [Online].

TRINH, K. T. 2010a. Logarithmic Correlations For Turbulent Pipe Flow Of Power Law Fluids. *arXiv.org [phys.fluid-dyn]* [Online]. Available: http://arxiv.org/abs/1007.0789.

TRINH, K. T. 2010b. A Zonal Similarity Analysis of Velocity Profiles in Wall-Bounded Turbulent Shear Flows. *arXiv.org 1001.1594 [phys.fluid-dyn]* [Online].

YOO, S. S. 1974. *Heat transfer and friction factors for non-Newtonian fluids in turbulent flow, PhD thesis,* US, University of Illinois at Chicago Circle.